%% file: eprint_dpf2013.tex
\pdfoutput=1
\documentclass[12pt]{article}
\usepackage{graphicx}
\usepackage{multicol}
\usepackage{bm}
\usepackage{amssymb}
\usepackage{amsmath}
\usepackage{cite}
\usepackage{caption}
\usepackage{subcaption}
\usepackage{mciteplus}
\usepackage{hyperref} 
\usepackage{bookmark}
\usepackage[all]{hypcap} 
\usepackage{ifthen} 
\newboolean{articletitles}
\setboolean{articletitles}{true} 

\newcommand\pubnumber{DPF2013-56}
\newcommand\pubdate{\today}

\def\napoli{Department of Physics\\University of Cincinnati, Cincinnati, Ohio 45221-0011 USA
}
\def\support{\footnote{On Behalf of the LHCb Collaboration}}

\def\Title#1{\begin{center} {\Large #1 } \end{center}}
\def\Author#1{\begin{center}{ \sc #1} \end{center}}
\def\Address#1{\begin{center}{ \it #1} \end{center}}

\newcommand\pubblock{\rightline{\begin{tabular}{l} \pubnumber\\
         \pubdate  \end{tabular}}}
\newenvironment{Abstract}{\begin{quotation}  }{\end{quotation}}
\newenvironment{Presented}{\begin{quotation} \begin{center} 
             PRESENTED AT\end{center}\bigskip 
      \begin{center}\begin{large}}{\end{large}\end{center} \end{quotation}}

\input econfmacros.tex
\begin{document}
\begin{titlepage}
\pubblock

\vfill
\Title{Studies of Charm Mixing and CPV}
\vfill
\Author{Adam C. Davis\support}
\Address{\napoli}
\vfill
\begin{Abstract}
LHCb has collected the world's largest sample of charmed hadrons.  This sample is used to search for direct and indirect CP violation in charm and to measure $D^0$ mixing parameters. Preliminary measurements from several decay modes are presented, with complementary time-dependent and time-integrated analyses.
\end{Abstract}
\vfill
\begin{Presented}
DPF 2013\\
The Meeting of the American Physical Society\\
Division of Particles and Fields\\
Santa Cruz, California, August 13--17, 2013\\
\end{Presented}
\vfill
\end{titlepage}
\def\thefootnote{\fnsymbol{footnote}}
\setcounter{footnote}{0}
\begin{flushleft}
\section{Introduction}

The neutral and charged charm meson sector is interesting for the study of Charge Parity (CP) violation in the Standard Model (SM) due to the presence of the $c$ quark. There are three types of CP violation which can be studied. When the magnitude of the amplitude for the decay of a $D$ meson into a certain final state $f$, $|\mathcal{A}_f|=|\langle f| H|D\rangle|$, does not equal the magnitude of the amplitude for the CP conjugate decay, $|\overline{\mathcal{A}}_{\overline{f}}| = |\langle\overline{f} | H |\overline{D}\rangle|$, there is direct CP violation. This can occur in both charged and uncharged $D$ meson decays. For neutral mesons, the misalignment of mass and flavor eigenstates causes spontaneous transitions between particle ($D^0$) and antiparticle ($\overline{D}^0$), and vice versa. This transition is known as flavor oscillation, or mixing. CP violation in mixing is then the asymmetry between oscillation from particle to antiparticle or vice versa, and finally there can be CP violation in the interference between mixing and decay. These proceedings report in summary the search for mixing in the neutral $D$ meson system, and searches for time integrated CP asymmetries in the decays $D^+\to\phi\pi^+$, $D_s^+\to K_S^0\pi^+$, $D^0\to K^-K^+\pi^+\pi^-$ and $D^0\to\pi^-\pi^+\pi^+\pi^-$ using 1 fb$^{-1}$ of data collected with the LHCb detector in 2011.

\section{Review of Theory}
It is customary to write the mass eigenstates in terms of the flavor eigenstates as $|D_{1,2}\rangle = p|D^0\rangle\pm q|\overline{D}^0\rangle$, and to characterize the mixing by the dimensionless parameters $x=(m_2-m_1)/\Gamma$ and $y = (\Gamma_2-\Gamma_1)/(2\Gamma)$, where $p$ and $q$ are complex numbers, $m_{1,2}$ and $\Gamma_{1,2}$ are the mass and decay width eigenvalues, and $\Gamma$ is the average width. If mixing did not occur, then $x=y=0$. In this formalism, CP violation in pure mixing occurs when $|q/p|\ne 1$. Finally, CP violation in the interference between mixing and decay depends on the quantity $\lambda_f\equiv\left(\frac{q}{p}\frac{\overline{\mathcal{A}}_f}{\mathcal{A}_f}\right)$ and $\lambda_{\overline{f}}\equiv\left(\frac{q}{p}\frac{\overline{\mathcal{A}}_{\overline{f}}}{\mathcal{A}_{\overline{f}}}\right)$.\\

Measurements in this system are interesting for a variety of reasons. First, the $D^0$ represents the only up-type quark system which undergoes mixing. In the Standard Model, short range interactions are dominated by GIM- and CKM-suppressed loop diagrams, and non-perturbative long range interactions may dominate these effects, making the theoretical calculations challenging. Standard model expectation is for $x$ and $y$ to be as large as $\mathcal{O}(10^{-2})$\cite{Falk1,Falk2}. In the case of direct CP violation, Standard Model predictions give CP asymmetries of $\mathcal{O}(10^{-3})$ and depends on final state. Indirect CP violation, specifically in mixing, is expected at $\mathcal{O}[|(V_{cb}V_{ub})/(V_{cs}V_{us})|] \sim 10^{-3}$. Any departure from this could point to new physics\cite{GrossmanNirKagan,Blaylock5,Petrov,Falk1,Falk2}.

\section{$D^0-\overline{D}^0$ Mixing}
In 2012, LHCb presented a search for mixing in the neutral $D$ system\cite{dmix_paper}. This was accomplished with the decay $D^{*+}\to D^0\pi_s^+, D^0\to K\pi$, where $\pi_s$ is the characteristic soft pion produced in the flavor conserving decay of the $D^*$. The $D^0$ can decay via ``right-sign" (RS), $D^0\to K^-\pi^+$ decay or the ``wrong-sign" (WS), $D^0\to K^+\pi^-$ decay. Unless noted otherwise, charge conjugate decays are implied. The RS (WS) signal yields are determined from a fit to the $D^0\pi_s^+$ invariant mass distributions, corresponding to $\approx 8.4$M ($\approx 36$k) events. The fits are shown in Figure~\ref{fig:yields}.

\begin{figure}[htb]
\begin{center}
\resizebox{0.7\textwidth}{!}{
	\begin{subfigure}[b]{0.435\textwidth}
                \includegraphics[width=\textwidth]{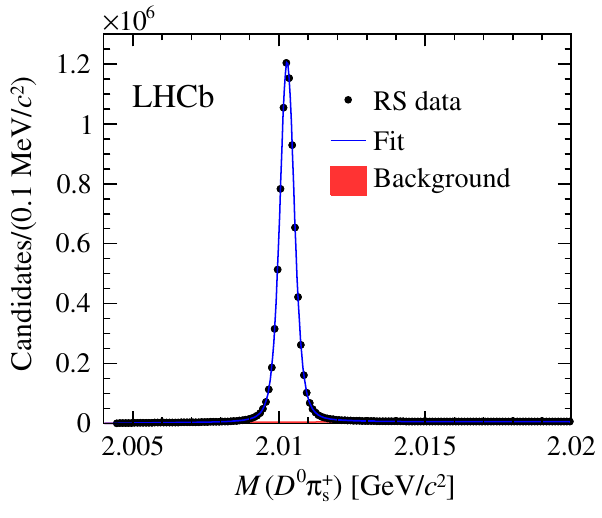}
                \caption{}
		\label{fig:RS_yield}
        \end{subfigure}\quad%
	\begin{subfigure}[b]{0.435\textwidth}
                \includegraphics[width=\textwidth]{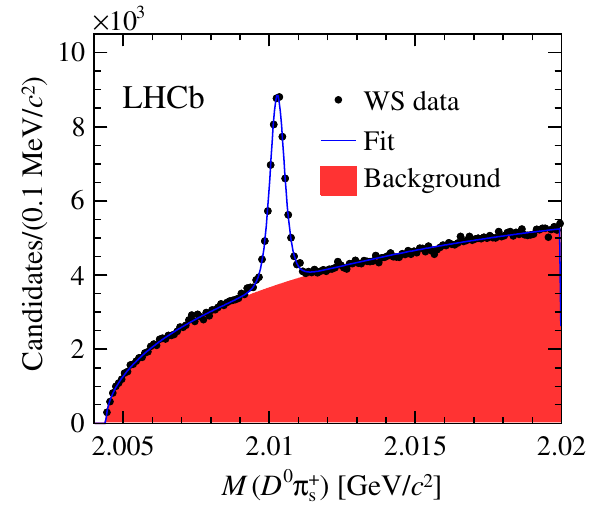}
                \caption{ }
		\label{fig:WS_yield}
        \end{subfigure}\quad%
        }
\caption{(a) RS yield fit. (b) WS yield fit. Black points represent the data, blue lines are the total fit models, and red indicates the background.}
\label{fig:yields}
\end{center}
\end{figure}

To a very good approximation, direct CPV can be ignored. RS decays are Cabibbo-favored (CF) and the decay rate approaches an exponential. WS decays, however, can either proceed via direct Doubly-Cabibbo-suppressed (DCS) decay, or first mix into a $\overline{D}^0$, then undergo a CF decay. Expanding in small $x$ and $y$, the ratio of decay rates then becomes

\begin{equation}
R(t) = \frac{N(WS(t))}{N(RS(t))} = R_D+\sqrt{R_D} y' (\Gamma t) + \frac{x'^2+y'^2}{4}(\Gamma t)^2,
\end{equation}

where $x'$ and $y'$ are $x$ and $y$ rotated by the strong phase difference between CF and DCS decays, 
\begin{equation*}
	\left(\begin{array}{c} x' \\ y' \end{array} \right)= \left( \begin{array}{c c} \cos\delta & \sin\delta \\ -\sin\delta & \cos\delta \end{array}\right) \left( \begin{array}{c} x \\ y \end{array} \right),\end{equation*}

and $R_D$ is the ratio of DCS to CF decay rates. If there were no mixing, then $x'=y'=0$, and $R(t)$ would be constant. By measuring this ratio as a function of time, it is possible to extract the mixing parameters $R_D$, $x'^2$, and $y'$. Figure~\ref{fig:R_D_plot} shows the reconstructed distribution, which disfavors the no-mixing hypothesis. Figure~\ref{fig:xy_contours_plot} shows the extracted contours for the mixing parameters $x'^2$ and $y'$, and show that no mixing is excluded at 9.1$\sigma$.  Finally, Figure~\ref{fig:comparison_contours_plot} shows the comparison of the LHCb result with other experiments\cite{PhysRevLett.96.151801, Aubert:2007wf, Aaltonen:2007ac}.

\begin{figure}[tbf]
\begin{center}
	\resizebox{\textwidth}{!}{
	\begin{subfigure}[b]{0.435\textwidth}
                \includegraphics[width=\textwidth]{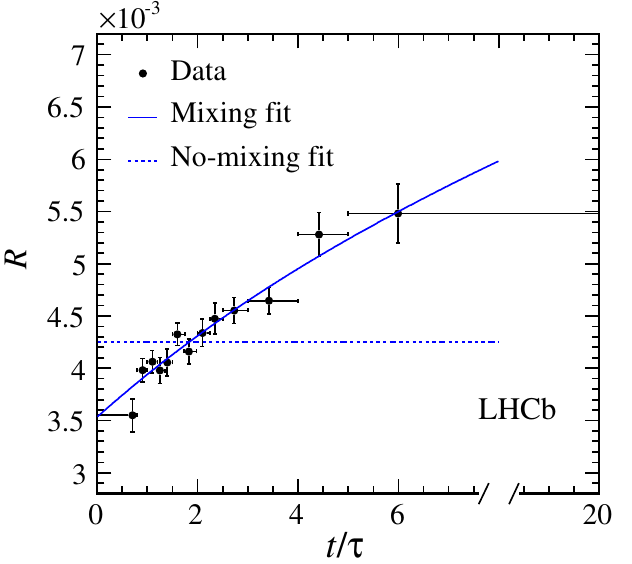}
                \caption{ }
		\label{fig:R_D_plot}
        \end{subfigure}\quad%
	\begin{subfigure}[b]{0.4\textwidth}
                \includegraphics[width=\textwidth]{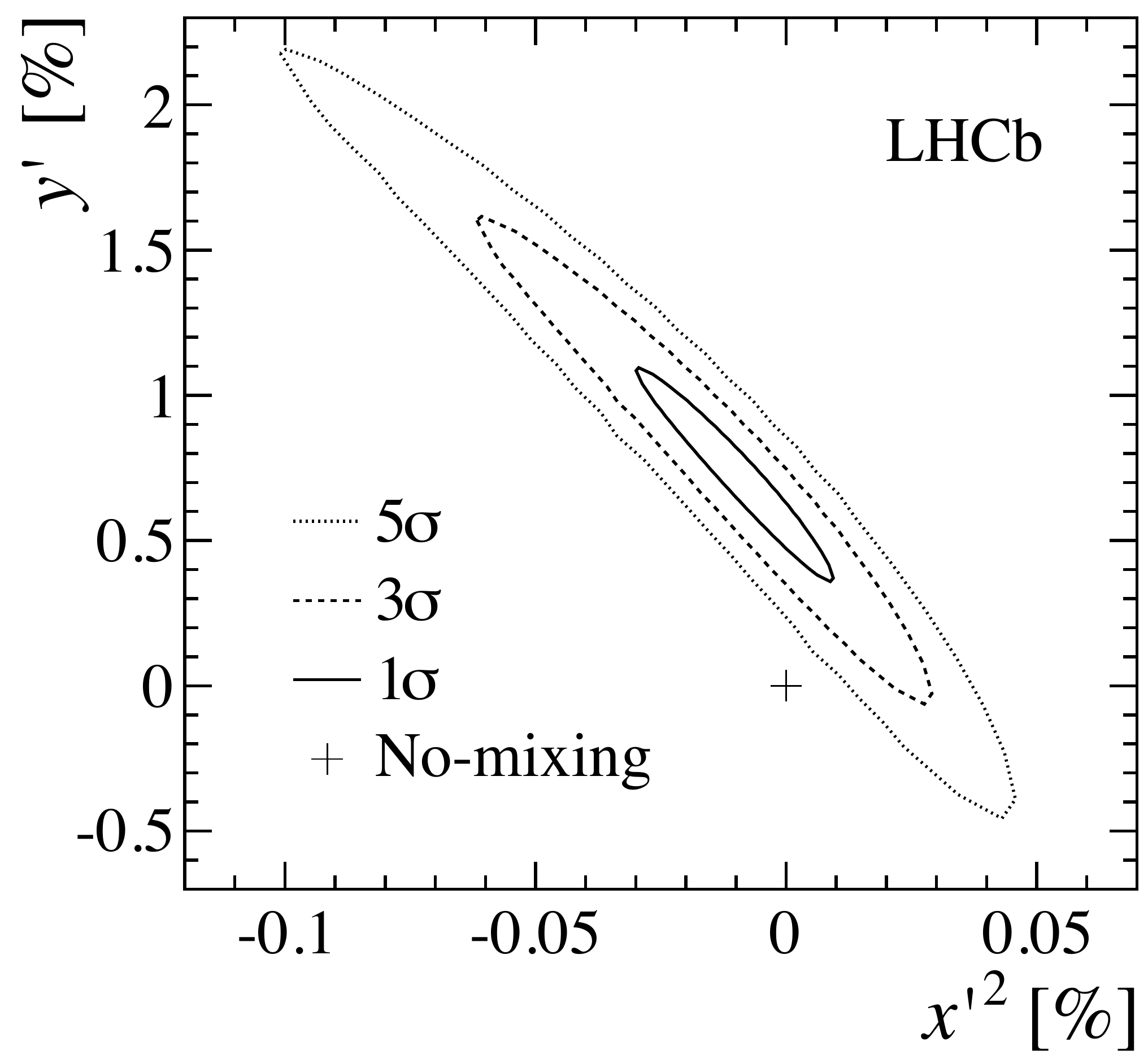}
                \caption{ }
		\label{fig:xy_contours_plot}
        \end{subfigure}\quad%
	\begin{subfigure}[b]{0.36\textwidth}
                \includegraphics[width=\textwidth]{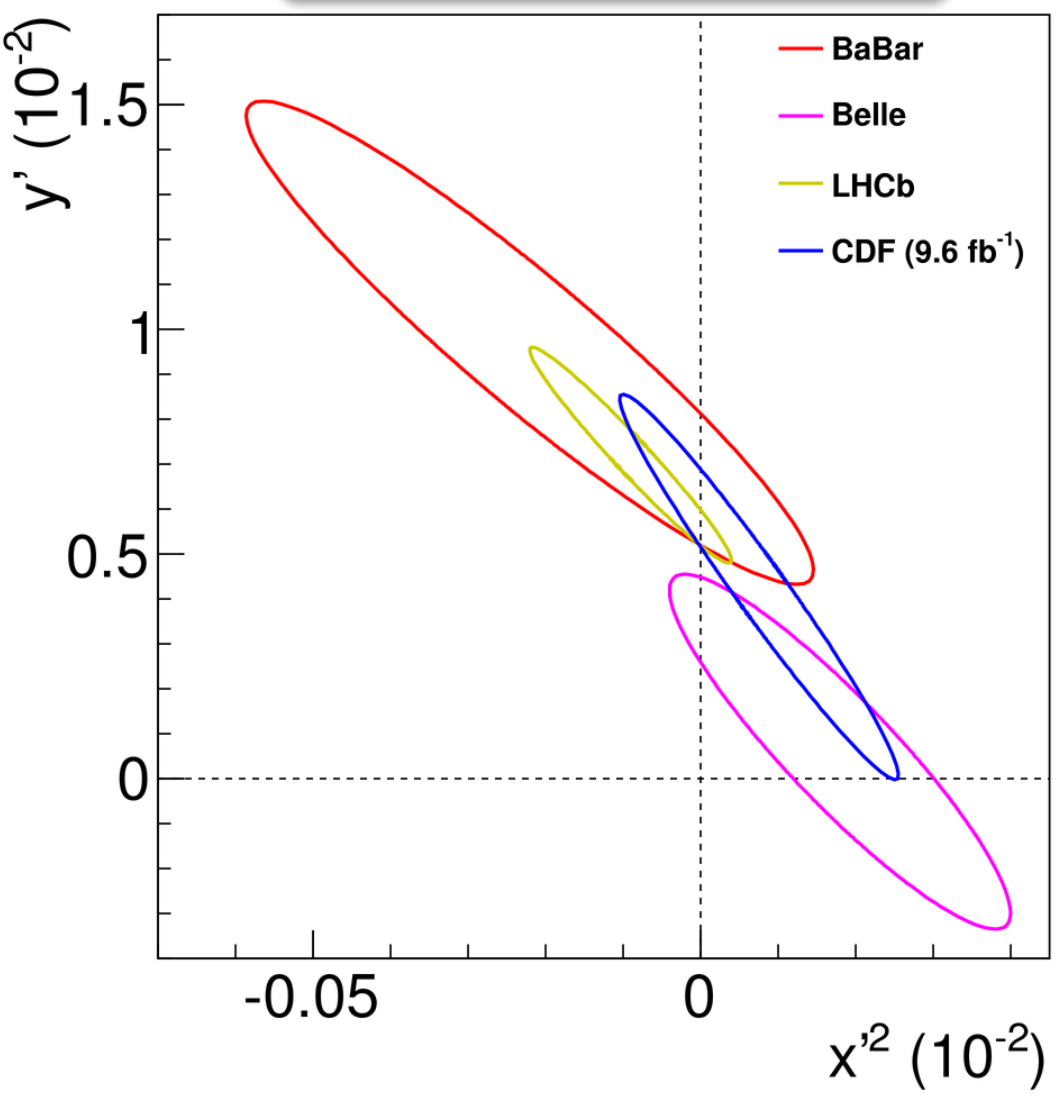}
                \caption{ }
		\label{fig:comparison_contours_plot}
        \end{subfigure}%	
        }
\label{fig:mixing_stuff}
\caption{(a) Time dependence of the WS/RS ratio, $R(t)$. The dashed line is the fit under the no mixing hypothesis. The solid line is the fit allowing for mixing hypothesis. (b) Two dimensional contours of extracted $x'^2$ and $y'$. (c) Comparison to other experiments.}
\end{center}
\end{figure}

\section{CP Violation in $D^+\to\phi\pi^+$ and $D_s^+\to K_S^0\pi^+$ }
In addition to mixing, LHCb carried out searches for direct CPV in the decays $D^+\to\phi\pi^+$ and $D_s^+\to K_S^0\pi^+$\cite{dphipi}. There can only be direct CP violation in charged mesons, and could arise from the interference between tree-level and penguin amplitudes. In order to measure the CP violation, two CP asymmetries can be written:

\begin{equation}\label{eq:dphipi}
	A_{CP} (D^+\to\phi\pi^+) = A_\text{raw}(D^+\to\phi\pi^+)-A_\text{raw}(D^+\to K_S^0\pi^+)+A_{CP}(K^0/\overline{K}^0)
\end{equation}
\begin{equation}\label{eq:dskspi}
	A_{CP}(D_s^+\to K_S^0\pi^+)=A_\text{raw}(D_s^+\to K_S^0\pi^+)-A_\text{raw}(D_s^+\to\phi\pi^+)+A_{CP}(K^0/\overline{K}^0).
\end{equation}

\begin{figure}[htb]
\begin{center}
\includegraphics[width=0.7\textwidth]{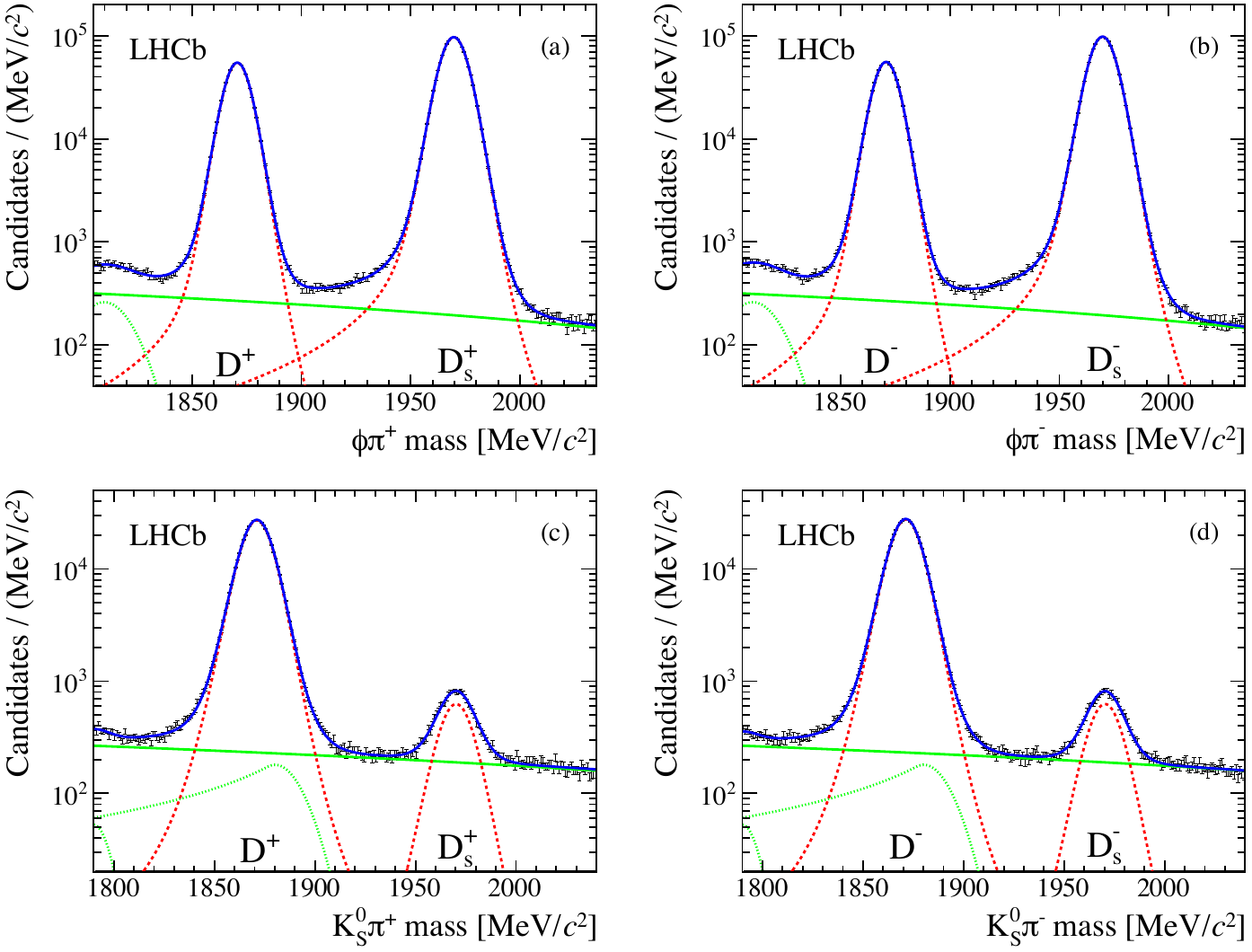}
\caption{Fits to $D^\pm\to\phi\pi^\pm$[(a) and (b)], and $D_s^\pm\to K_S^0\pi^\pm$ [(c) and (d)] invariant mass distributions. (a) and (c) are fits to the $D_{(s)}^+$, (b) and (d) are fits to the $D_{(s)}^-$ distributions. In each case, the points represent the data, the solid blue line is the total fit model, dashed red are the signal components, solid green is the combinatorial background, and dashed green represents misreconstructed decays.}
\label{fig:dphi_fits}
\end{center}
\end{figure}

The raw asymmetries above, $A_\text{raw}= \left(N_{D_{(s)}^+}-N_{D_{(s)}^-}\right)/\left(N_{D_{(s)}^+}+N_{D_{(s)}^-}\right)$, are taken from data, and $A_{CP}(K^0/\overline{K}^0)$ is the correction for CP violation in the neutral kaon system, as discussed in \cite{LHCb:2012dpmasymm}. The terms $N_{D_{(s)}^{\pm}}$ are the yields for the $D_{(s)}^\pm$ decays. The middle term in Equations~\ref{eq:dphipi} and~\ref{eq:dskspi} has negligible CP violation in both cases, and is present to cancel production and detector asymmetries. The yields are extracted from a fit to the $\phi\pi^\pm$ and $K_S^0\pi^\pm$ invariant mass distributions, shown in Figure~\ref{fig:dphi_fits}. The final asymmetries are given by

\begin{align}
A_{CP} (D^+\to\phi\pi^+) & = (-0.04\pm0.14\pm0.14)\%,\\
A_{CP} (D_s^+\to K_S^0\pi^+) & = (+0.61\pm 0.83\pm0.14)\%.
\end{align}

Both results are consistent with CP conservation. It is possible, however, that due to the variation of the strong phase across the Dalitz plot, all CP violation could be cancelled out. This variation is illustrated in Figure~\ref{fig:mc_dalitz}. In order to account for this variation, a new observable, 
\begin{equation}
A_{CP|S}\equiv\frac{1}{2}(A_{CP}^A+A_{CP}^C-A_{CP}^B-A_{CP}^D),
\end{equation}

is defined, where $A_{CP}^X$ defines the CP asymmetry in region X of the Dalitz plot, as assigned in Figure~\ref{fig:dalitz}. This observable is sensitive to CP asymmetries which would be cancelled in $A_{CP}$. The distribution of events across the Dalitz plot is shown in Figure~\ref{fig:rd_dalitz}, and the result is

\begin{equation}
A_{CP|S}=(-0.18\pm0.17\pm0.18)\%.
\end{equation}

This is consistent with CP conservation.
\begin{figure}[htb]
\begin{center}
\resizebox{0.95\textwidth}{!}{
	\begin{subfigure}[b]{0.5\textwidth}
		\includegraphics[width=\textwidth]{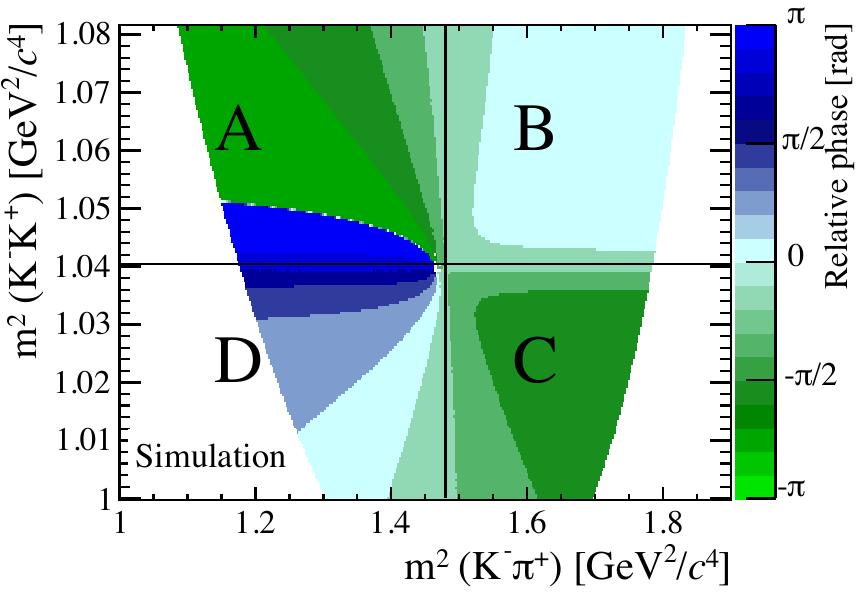}               
                \caption{ }
		\label{fig:mc_dalitz}
        \end{subfigure}\quad%
	\begin{subfigure}[b]{0.5\textwidth}
		\includegraphics[width=\textwidth]{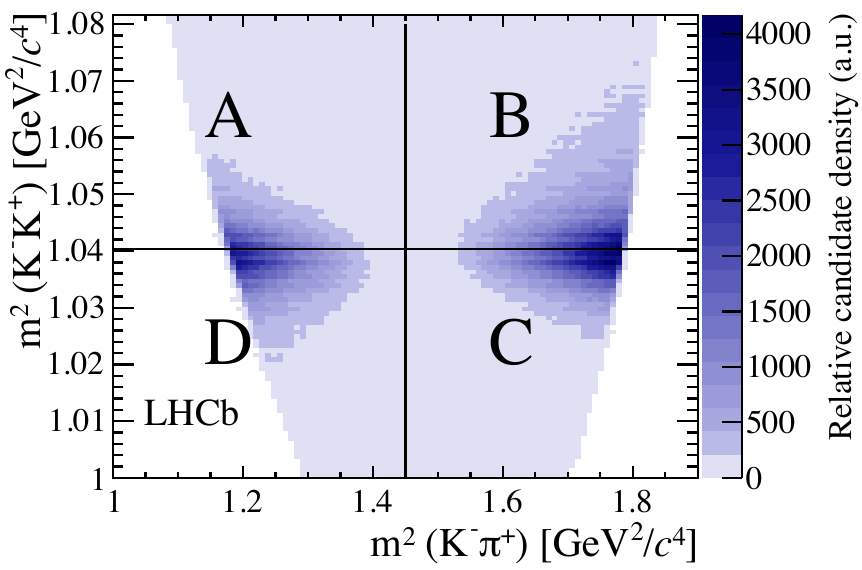}
                \caption{ }
		\label{fig:rd_dalitz}
        \end{subfigure}
        }
\caption{(a) Simulation of the strong phase variation over the $\phi$ region of the $D^+$ Dalitz plot. The regions A, B, C and D are defined for the new observable, $A_{CP|S}$. (b) Observed events in the $D^+\to \phi \pi^+$ Dalitz plot, with regions overlayed.}
\label{fig:dalitz}
\end{center}
\end{figure}

\section{CP Violation in $D^0\to K^-K^+\pi^+\pi^-$ and $D^0\to \pi^-\pi^+\pi^+\pi^-$}

The final analysis presented is a model-independent search for CP violation in the decay of $D^0\to K^-K^+\pi^-\pi^+$ and $D^0\to \pi^-\pi^+\pi^+\pi^-$\cite{d4h}. In these decays, CP violation could manifest itself in the interference between tree-level and penguin amplitudes. The search for CP violation is performed over the entire 4-body phase space of each decay. To fully describe this phase-space, five invariant mass combinations are needed. As an example, one combination which could describe the entire decay is 

\begin{equation}
D^0\to(1,2,3,4): \{s(1,2),s(2,3),s(3,4),s(1,2,3),s(2,3,4)\},
\end{equation}

where 1-4 are the decay products and $s$ is the invariant mass squared of the combinations of particles given. In this analysis, identical particles are assigned randomly. The channel $D^0\to K^-\pi^+\pi^+\pi^-$ is used as a control channel. \\
Signal candidates are selected through the decay $D^{*+}\to D^0\pi_s^+$. To determine the asymmetries, signal distributions are extracted via the $_s\mathcal{P}lot$ technique\cite{Pivk:2004ty}. The corresponding $_s\mathcal{W}$eights are extracted by fitting the $(m,\Delta m)$ plane, where $m$ is the mass of the $D^0$ candidate and $\Delta m$ is the mass difference between the $D^{*+}$and $D^0$ candidate. Projections of these fits are given in Figure~\ref{fig:splots}.

\begin{figure}[htb]
\begin{center}
\includegraphics[width=0.3\textwidth]{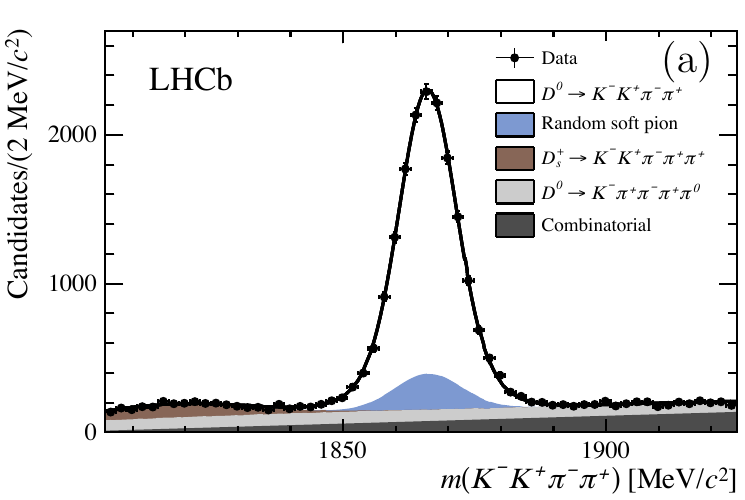}\includegraphics[width=0.3\textwidth]{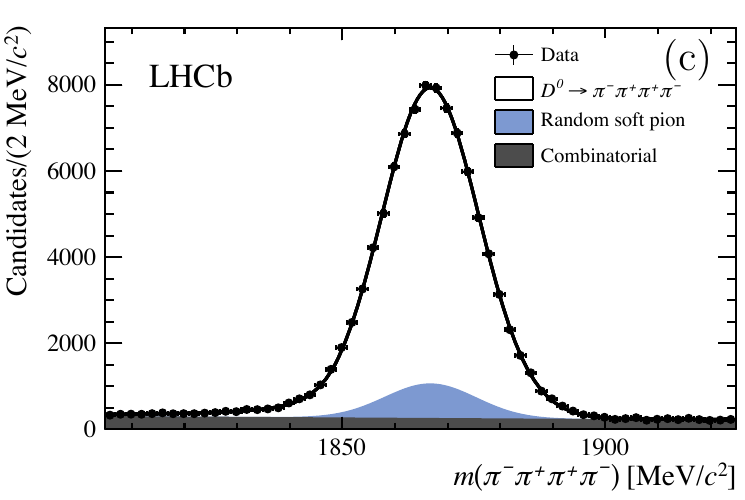}\includegraphics[width=0.3\textwidth]{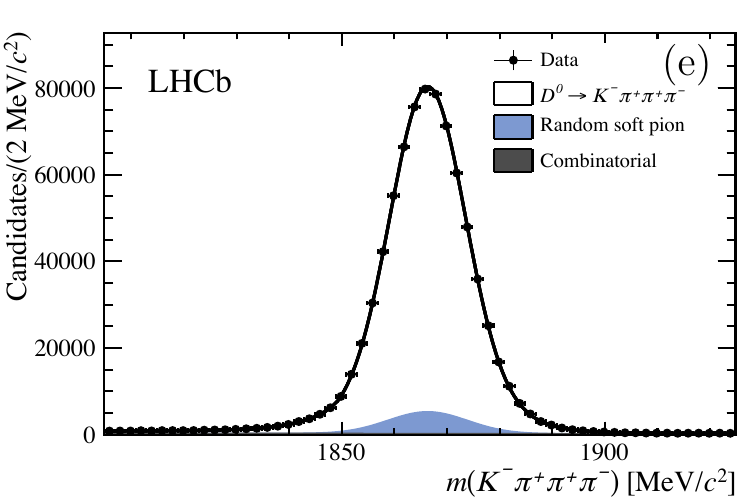}\\
\includegraphics[width=0.3\textwidth]{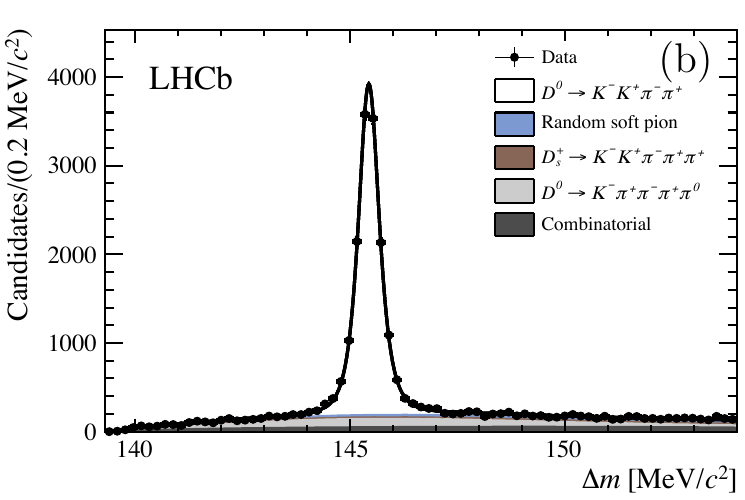}\includegraphics[width=0.3\textwidth]{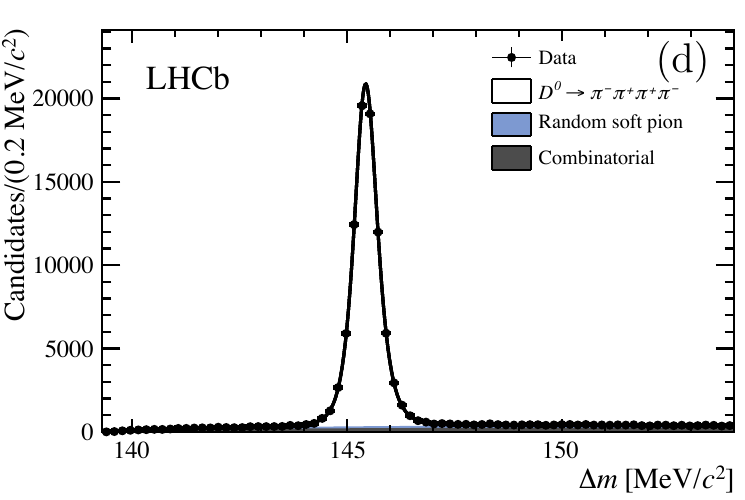}\includegraphics[width=0.3\textwidth]{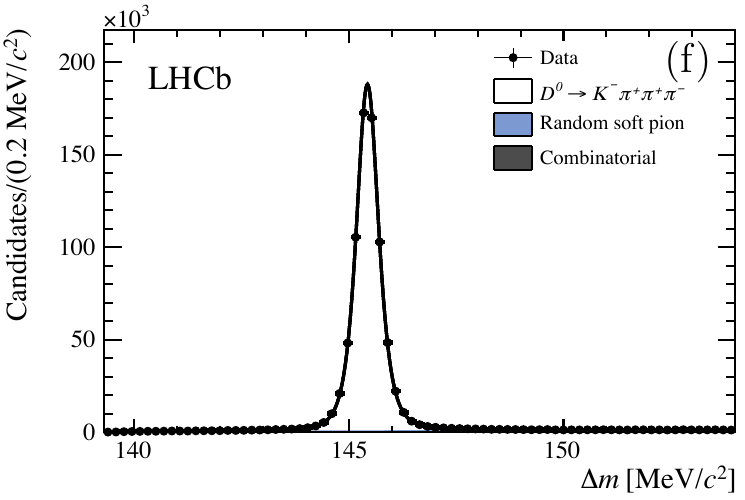}
\caption{Projections of fits to the $(m,\Delta m)$ plane of the decays $D^0\to K^-K^+\pi^+\pi^-$ [(a)-(b)], $D^0\to\pi^-\pi^+\pi^+\pi^-$ [(c)-(d)], and $D^0\to K^-\pi^+\pi^+\pi^-$ [(e)-(f)]. Points represent the data, the solid black line is the fit, and the shaded areas represent the background contributions. The corresponding signal yields are $330,000$ $D^0\to K^-K^+\pi^+\pi^-$ decays, 57,000 $D^0\to\pi^-\pi^+\pi^+\pi^-$ decays, and $2.9\times10^6$ $D^0\to K^-\pi^+\pi^+\pi^-$ decays.}
\label{fig:splots}
\end{center}
\end{figure}
The significance is then calculated in equally populated bins of 5D phase space as 

\begin{equation}
	S_{CP}^i=\frac{N_i(D^0)-\alpha N_i(\overline{D}^0)}{\sqrt{\alpha (\sigma_i^2(D^0)+\sigma_i^2(\overline{D}^0))}},
\end{equation}
where $N_i(D^0(\overline{D}^0))$ are the yields in of $D^0(\overline{D}^0)$ in bin $i$, $\sigma_i$ is the associated uncertainty, and the factor $\alpha=\frac{\sum_i N_i(D^0)}{\sum_i N_i(\overline{D}^0)}$, corrects for any global production asymmetries between $D^{*+}$ and $D^{*-}$. When summed over all bins $i$, the significance approaches a $\chi^2$ distribution with $N_\text{bins}-1$ degrees of freedom, from which a $p$-value can be calculated. In the case of CP conservation, the distribution of $S_{CP}^i$ is Gaussian.\\
The final results are given in Figure~\ref{fig:scparaw} and summarized in Table~\ref{table:thetab}. The raw CP asymmetry, $A_\text{raw}$ distributions are shown for comparison to $S_{CP}$ and do not indicate a CP asymmetry. Results are consistent with CP conservation in $D^0\to K^-K^+\pi^+\pi^-$ and $D^0\to \pi^-\pi^+\pi^+\pi^-$ decays, and no local asymmetries are found in the control channel.

\begin{figure}[htb]
\begin{center}
\includegraphics[width=0.3\textwidth]{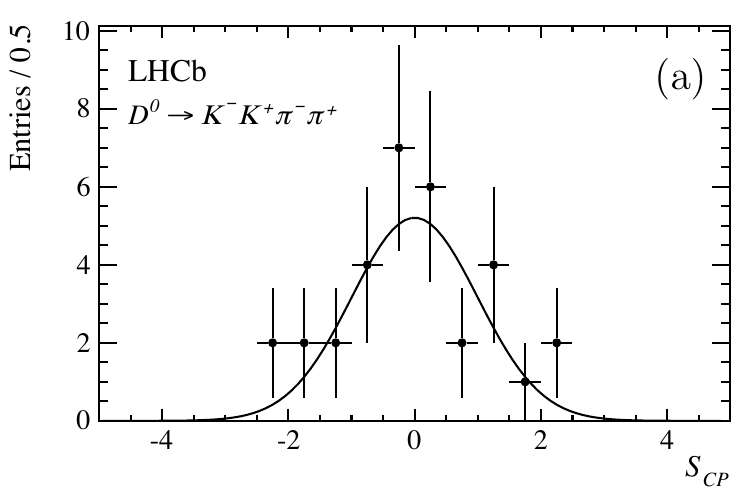}\includegraphics[width=0.3\textwidth]{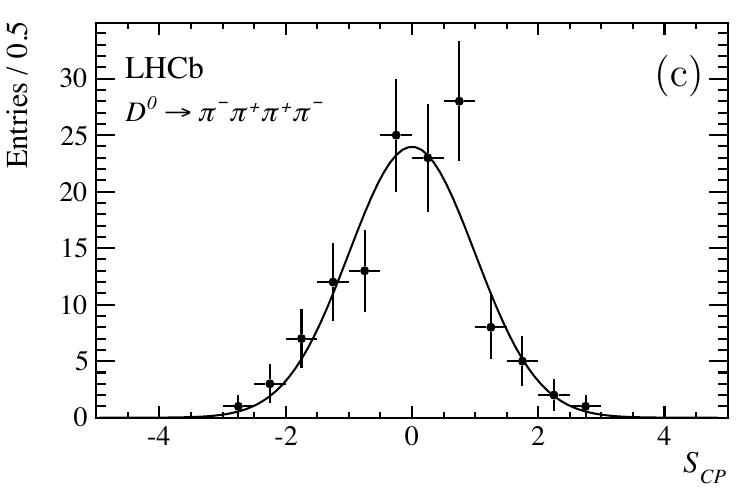}\includegraphics[width=0.3\textwidth]{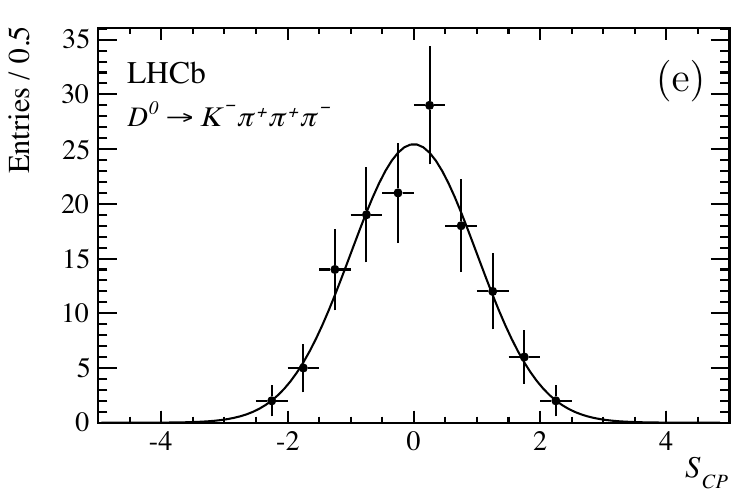}\\
\includegraphics[width=0.3\textwidth]{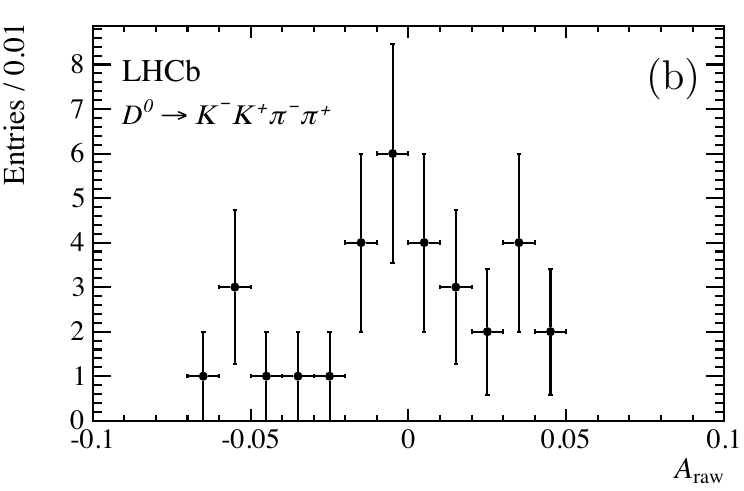}\includegraphics[width=0.3\textwidth]{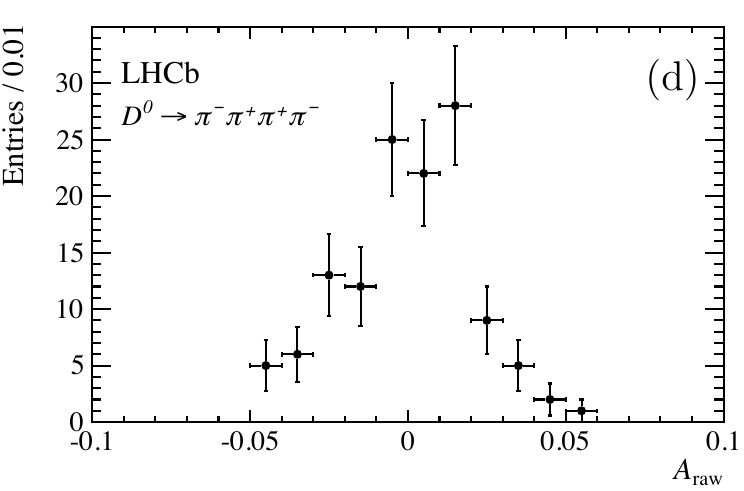}\includegraphics[width=0.3\textwidth]{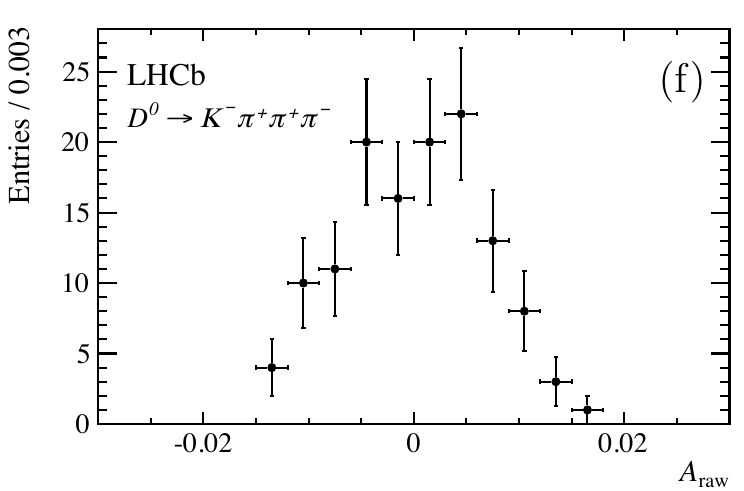}
\caption{Distributions of $S_{CP}$[(a),(c) and (e)] and $A_\text{raw}$[(b),(d) and (f)] for $D^0\to K^-K^+\pi^-\pi^+$[(a),(b)], $D^0\to \pi^-\pi^+\pi^+\pi^-$ [(c),(d)] and $D^0\to K^-\pi^+\pi^+\pi^-$[(e),(f)]. Plots correspond to the middle row of Table~\ref{table:thetab}.}
\label{fig:scparaw}
\end{center}
\end{figure}
\begin{table}[t]
\caption{Results of fit to 5D phase space of (a) $D^0\to K^-K^+\pi^+\pi^-$, (b) $D^0\to \pi^-\pi^+\pi^+\pi^-$, and  (c) $D^0\to K^-\pi^+\pi^+\pi^-$ in different binnings. Results are consistent with CP conservation.}
\begin{center}
\resizebox{0.98\textwidth}{!}{
\subfloat[]{
\begin{tabular}{l|ccc}
\hline
\multicolumn{3}{c}{$D^0\to K^-K^+\pi^+\pi^-$}\\ \hline
	Bins & $\chi^2/\text{ndf}$ & $p-$value(\%) \\ \hline \hline
	16 & 22.7/15 & 9.1 \\ \hline
	32 & 42.0/31 & 9.1 \\ \hline
	64 & 75.7/63 & 13.1 \\ \hline
\end{tabular}
}~
\subfloat[]{
\begin{tabular}{l|ccc}
\hline
\multicolumn{3}{c}{$D^0\to\pi^-\pi^+\pi^+\pi^-$}\\ \hline
	Bins & $\chi^2/\text{ndf}$ & $p-$value(\%) \\ \hline \hline
	64 & 68.8/63 & 28.8 \\ \hline
	128 &130.0/127 & 41.0 \\ \hline
	256 & 247.7/255 & 61.7 \\ \hline
\end{tabular}
}~
\subfloat[]{
\begin{tabular}{l|ccc}
\hline
\multicolumn{3}{c}{$D^0\to K^-\pi^+\pi^+\pi^-$}\\ \hline
	Bins & $\chi^2/\text{ndf}$ & $p-$value(\%) \\ \hline \hline
	16 & 16.5/15 & 34.8 \\ \hline
	128 & 113.4/127 & 80.0 \\ \hline
	1024 & 1057.5/1023 & 22.1 \\ \hline
\end{tabular}
}
}
\end{center}
\label{table:thetab}
\end{table}%

\section{Conclusion}
With 1 fb$^{-1}$ of data collected in 2011 by the LHCb detector, $D^0-\overline{D}^0$ mixing has been verified, and strong constraints on CPV in the decays $D^+\to\phi\pi^+$, $D_s^+\to K_S^0\pi^+$, $D^0\to\pi^-\pi^+\pi^+\pi^-$ and $D^0\to K^-K^+\pi^+\pi^-$ are reported. With an additional 2 fb$^{-1}$ of data currently being analyzed, many more results will come soon.
%\end{linenumbers}
\end{flushleft}
\bibliographystyle{LHCb}
\bibliography{eprint_dpf2013}

\end{document}

%% file: econfmacros.tex
\def\beq{\begin{equation}}
\def\eeq#1{\label{#1}\end{equation}}
\def\eeqn{\end{equation}}

\def\beqa{\begin{eqnarray}}
\def\eeqa#1{\label{#1}\end{eqnarray}}
\def\eeqan{\end{eqnarray}}

\let\bar=\overbar

\def\Dslash{\not{\hbox{\kern-4pt $D$}}}
\def\dslash{\not{\hbox{\kern-2pt $\del$}}}

\def\msb{{\bar{\ssstyle M \kern -1pt S}}}